\documentclass[aps,prd,preprint,nofootinbib,groupedaddress]{revtex4-1}

\usepackage{amsfonts,amssymb,amsmath}
\usepackage{graphicx}
\usepackage{color}
\usepackage{hyperref}

\usepackage{todonotes}

\usepackage[framemethod=TikZ]{mdframed}
\usepackage{lipsum}

\mdfdefinestyle{MyFrame}{
linecolor=blue,
outerlinewidth=2pt,
roundcorner=20pt,
innertopmargin=\baselineskip,
innerbottommargin=\baselineskip,
innerrightmargin=20pt,
innerleftmargin=20pt,
backgroundcolor=gray!20!white}

\newcommand{\be}{\begin{equation}}
\newcommand{\ee}{\end{equation}}
\newcommand{\bea}{\begin{eqnarray}}
\newcommand{\eea}{\end{eqnarray}}
\newcommand{\beal}{\begin{aligned}}
\newcommand{\eeal}{\end{aligned}}

\usepackage{color}

\def\ub{U_{b}}
\def\vb{V_{b}}
\def\uc{U_{c}}
\def\vc{V_{c}}
\def\df{\dot{f}}

\def\deta{\dot{\eta}}

\def\dps{\dot{\phi}^2 }
\def\fsds{ f_{0} }

\def\be{\begin{equation}}
\def\ee{\end{equation}}
\def\bea{\begin{eqnarray}}
\def\eea{\end{eqnarray}}
\def\ba{\begin{align}}
\def\ea{\end{align}}

\begin{document}

\preprint{ACFI-T18-05}
\preprint{DCPT-18/13}

\title{Evolving Black Holes in Inflation}

\author{Ruth Gregory${}^{a,b}$}
\email{r.a.w.gregory@durham.ac.uk }

\author{David Kastor${}^c$}
\email{kastor@physics.umass.edu}

\author{Jennie Traschen${}^c$}
\email{traschen@physics.umass.edu}

\affiliation{${}^a$Centre for Particle Theory, Durham University,
South Road, Durham, DH1 3LE, UK\\
${}^b$Perimeter Institute, 31 Caroline St, Waterloo, Ontario N2L 2Y5,
Canada\\
${}^c$Amherst Center for Fundamental Interactions, Department of Physics,
University of Massachusetts, 710 N Pleasant St, Amherst, MA 01003, USA}

\date{\today}

~\\

\vskip 1cm

\begin{abstract}

We present an analytic, perturbative solution to the Einstein equations
with a scalar field that describes dynamical black holes in a slow-roll 
inflationary cosmology. We show that the metric evolves quasi-statically 
through a sequence of Schwarzschild-de Sitter like metrics with time 
dependent cosmological constant and mass parameters, such that the 
cosmological constant is instantaneously equal to the value of the scalar potential. 
The areas of the black hole and cosmological horizons each increase in 
time as the effective cosmological constant decreases, and the fractional 
area increase  is proportional to the fractional change of the cosmological 
constant, times a geometrical factor. For black holes ranging in size from 
much smaller than to comparable to the cosmological horizon, the 
pre-factor varies from very small to order one. The ``mass first law" and 
the ``Schwarzchild-de Sitter patch first law" of thermodynamics
are satisfied throughout the evolution.

\end{abstract}

\pacs{PACS}

\maketitle

\section{Introduction}

The dynamics of black holes in cosmology is important for understanding 
several questions about the very early universe. 
During very hot phases a population of black holes will affect the 
behavior of the plasma and impact upon phase transitions, and could 
even have a significant effect on the the geometry near the big-bang. 
On the late time, or `cooler' side, there has been interesting recent research 
\cite{Bird:2016dcv,Sasaki:2016jop,Carr:2016drx,Sasaki:2018dmp}
about the possibility of an early population of black holes that could seed 
galaxies and provide progenitors for the larger black holes detected by 
LIGO \cite{Abbott:2016blz,TheLIGOScientific:2016htt}.

Such situations are highly interactive involving classical accretion, 
the expansion of the universe, as well as classical and quantum mechanical radiation
exchange. Known analytic solutions include the stationary 
Kerr-Reissner-Nordstrom de Sitter metric \cite{Carter:1973rla},
the cosmological McVittie black hole spacetime \cite{McVittie:1933zz}, 
which has an unphysical pressure field except when it reduces to 
Schwarzschild-de Sitter \cite{Kaloper:2010ec}, and cases of
multi, maximally charged, black holes
\cite{Kastor:1992nn,Horne:1993sy,Gibbons:2009dr,Maeda:2010aj} 
that exploit fake supersymmetries
\cite{Chimento:2012mg,Klemm:2015qpi}.
A range of approximations and numerical analyses have been used 
to study accretion and estimate the 
growth of black holes in these interactive systems, including 
\cite{Frolov:2002va,Jacobson:1999vr,Saida:2000at,Harada:2004pf,
Sultana:2005tp,MartinMoruno:2006mi,JimenezMadrid:2005rk,
MartinMoruno:2008vy,Faraoni:2007es,Carrera:2008pi,Rodrigues:2009eg,
Carr:2010wk,UrenaLopez:2011fd,Guariento:2012ri,Rodrigues:2012xm,
Chadburn:2013mta,Abdalla:2013ara,Babichev:2005py,Babichev:2008jb,
Babichev:2012sg,Babichev:2014lda,Davis:2014tea,
Afshordi:2014qaa,Davis:2016avf}. 
Many of these studies incorporate test stress-energy and infer 
time rate of change of the black hole mass by computing the flux of 
stress-energy across the horizon.
In this paper we continue our studies of the evolution of black holes in 
inflationary cosmologies  by solving the full coupled scalar field plus Einstein
equations in the slow-roll and perturbative approximations.
This system has the advantage that it is only ``mildly dynamical",
and quantities can be computed in controlled approximations. While mild, 
we anticipate that the solution presented here will be useful for computing 
observational signals of black holes that are present during the inflationary 
epoch, and helpful in developing techniques applicable to hotter early 
universe situations.

Slow-roll inflationary cosmologies are described by quasi-de Sitter metrics, 
in which the cosmological constant, $\Lambda$, is provided by the potential 
of a scalar field $\phi$, and $\Lambda$ slowly evolves in time
as $\phi$ rolls down a potential. If a black hole is included, one thinks 
about the spacetime as being approximately described by 
quasi-Schwarzschild-de Sitter metrics,
with both $\Lambda$ and the mass parameter $M$ changing slowly. 
But is this picture correct? Here we answer this question in the affirmative.
We show that the black hole plus scalar field system evolves through a 
sequence of very nearly
Schwarzschild-de Sitter (SdS) metrics as the inflaton rolls down the potential. 
 
In earlier work, Chadburn and Gregory \cite{Chadburn:2013mta} found 
perturbative solutions for the black hole and scalar field cosmology, 
as the field evolves slowly in an exponential potential. 
Their null-coordinate techniques were particularly useful for analyzing 
the behavior of the horizon, and
computing the growth of the horizon areas. 
Recently we extended the work of \cite{Chadburn:2013mta}
to general potentials \cite{Gregory:2017sor}, with a focus on evolutions 
that interpolate between an initial SdS and final SdS with a smaller cosmological
constant $\Lambda$. The analysis yielded geometrical expressions for 
the total change in the horizon areas. Further, it was found that the ``SdS-patch" 
first law was obeyed between the initial and final SdS states. 
The SdS-patch first law \cite{Dolan:2013ft} only involves quantities defined in
the portion of the spacetime between the black hole and cosmological horizons, 
see equation \eqref{firstsds} below. That result supports the picture of
a quasi-static evolution through SdS metrics, which we address in detail in this paper.

Consider Einstein gravity coupled to a scalar field with potential 
$W(\phi)$ governed by the action
\be\label{action}
S=\int d^4x\sqrt{-g}\left(\frac{M_p^2}{2}R-(\nabla\phi)^2-2W(\phi)\right)
\ee
where $M_p^2 = 1/8\pi G$ is the reduced Planck Mass in units with
$\hbar=c=1$.
Varying the action yields Einstein's equation $G_{ab}=T_{ab}/M_p^2$ 
with the stress-energy tensor  given by
\be\label{stress}
T_{ab} = \nabla_a\phi\nabla_b\phi 
-\frac12 g_{ab}\left [(\nabla\phi)^2+2W(\phi)\right]
\ee
and the scalar field equation of motion is given by
\be
g^{ab}\nabla_a \nabla_b \phi = \frac{\partial W}{\partial\phi}\,.
\label{scalar}
\ee

The coordinate system used in \cite{Gregory:2017sor}
was chosen to facilitate analysis of the horizons and clarify the
degrees of freedom of the metric, but did not prove 
amenable for finding the full time-dependent corrections to the metric 
throughout the SdS patch. In this paper we aim to find the metric in a 
transparent and physically useful form. In particular, the metric ansatz 
ultimately promotes the SdS parameters $\Lambda$ and $M$ to time 
dependent functions $\Lambda (T)$ and $M(T)$ in a natural way. 
The solution determines $d\Lambda / dT$ and $dM/dT$, which are 
found to be proportional to $(d\phi /dT )^2$. Additional analysis gives the rate of 
change of the area of the black hole horizon $A_b$ to be
\be\label{dotareabh}
\frac{dA_b}{dT} = \frac{A_b }{\kappa_b M_p^2} \left(\frac{d\phi }{dT} \right)^2 
\ee
where $\kappa_b$ is the surface gravity of the background SdS black hole,
and $T$ a suitable time coordinate to be identified. 
A similar relation holds for the cosmological horizon, which also grows in time.
In the slow-roll approximation, $d\phi /dT$ is proportional to 
$\partial W/ \partial \phi$, so the complete solution is known once the 
potential and the initial black hole area are specified.
It is then shown that both the SdS-patch first law,  and the more familiar mass first law,
are obeyed throughout the evolution.

The analysis starts by asking: \emph{``Who sees the simple evolution?''}
In Section \ref{sdssection} we review basics of the static SdS spacetime, 
and then take some care in choosing a natural time coordinate $T$ 
such that for a slowly rolling scalar, $\phi$ only depends on $T$.  
Transforming SdS to the new coordinate system then facilitates finding a 
useful anzatz for the time dependent metric. 
The necessary conditions on the potential for this
approximation to be valid are found in \S \ref{slowroll}. 
In Section \ref{solutionsection} the linearized Einstein equations are  
solved. Section \ref{horsection} analyzes
the geometry near the horizons. In Section \ref{horizonsandfirst} the rate of change of the 
horizon areas is found, both the SdS-patch first law and the mass first law are shown
to hold throughout the evolution, and a candidate definition of the 
dynamical surface gravity
is computed. Conclusions and open questions are presented in Section \ref{conclusion}.

\section{Schwarzschild-de Sitter with a slowly evolving inflaton}
\label{sdssection}

Finding the metric and scalar field of a black hole in general in an inflationary cosmology
is a complicated dynamical problem, even in the spherically symmetric case. 
On the other hand, in the case that the inflaton and the effective cosmological 
constant change slowly, there is an expectation that the metric
evolves quasi-statically through a sequence of SdS-like metrics, with
$\Lambda$ approximately equal to the value of the potential at that time.
In this paper we address this expectation, and show that the picture is 
remarkably accurate within the slow-roll and perturbative approximations,
with some small adjustments. 
To be concrete, we assume that the scalar potential $W(\phi )$ has a 
maximum where the field starts, and a minimum to which it evolves, so that 
the metric is initially SdS with an initial value of the black hole horizon area, then
evolves to an SdS with different values of $\Lambda$ and the black hole area.

The first step is to come up with a workable ansatz for the metric that is 
compatible with a quasi-SdS evolution. The key question is then 
\emph{to whom} does the dynamical spacetime look simple? 
Analytically, the equations are likely to be simpler using a time coordinate 
$T$ in which the scalar field only depends on $T$. This would imply that 
the potential $W(\phi )$ also only depends on $T$, and hence is consistent 
with the idea that the potential acts as a slowly changing cosmological constant 
with $W(\phi )\simeq \Lambda (T)$. In the initial unstable SdS phase $\phi$ 
is a constant, so a time-dependent $\phi$ would then be first order in a 
perturbative expansion. Hence when solving the wave equation for $\phi (T)$, 
the metric takes its background or zeroth order values, which allows us to find 
the preferred coordinate $T$ without knowing the back-reacted metric.
 
In this section we review some required properties of SdS metrics,
then analyze the slow-roll wave equation in SdS to find $T$. We then
transform SdS to this new coordinate system, and base our ansatz for 
the back-reacted metric on this new slicing of SdS, as we know it is 
compatible with an evolution of $\phi$ that only depends on $T$. 
Finally, we identify the conditions for an analog of the slow roll 
approximation to be valid.

\subsection{Schwarzschild-de Sitter spacetime}

The starting point for our construction is Schwarzschild-de Sitter (SdS) spacetime
\begin{equation}\label{sds}
ds^2 = -\fsds (r) dt^2 +\frac{dr^2 }{\fsds (r) } +r^2 d\Omega^2,\qquad \fsds (r)  
=  1 -\frac{2GM}{r}  -\frac{\Lambda }{3}  r^2
\end{equation}
where $M$ is interpreted as the mass, and $\Lambda$ is a positive 
cosmological constant.  The coupled Einstein-scalar field system \eqref{action} 
will have SdS solutions if the potential $W(\phi)$ has a stationary point $\phi_0$, 
such that  $W(\phi_0)>0$. SdS solutions then exist having $\phi=\phi_0$ 
everywhere and $\Lambda=W(\phi_0)/M_p^2$.

The SdS metric is asymptotically de Sitter at large spatial distances.  
For $GM\le \sqrt{\Lambda}/3$, the SdS metric function $f_0(r)$ has three 
real roots, which we will label as $r_c$, $r_b$ and $r_n$, and the SdS 
metric function can be rewritten as 
\be
f_0(r) = - \frac{ \Lambda }{ 3r}(r- r_c ) (r-r_b ) (r-r_n ) 
\ee
The absence of a term linear in $r$ in $f_0(r)$ implies that  $r_n=-(r_b+r_c)$, 
and we will assume that $r_c\ge r_b\ge 0$.  The parameters $M$ and 
$\Lambda$ of the SdS metric are then related to these roots according to
\be\label{mlambda}
GM =\frac{r_c r_b (r_b + r_c ) }{2(r_b^2 +r_c^2 +r_b r_c )} 
\  , \quad  \Lambda =\frac{3}{r_b^2 +r_c^2 +r_b r_c}
\ee

Provided also that $M>0$, the SdS metric describes a black hole in de Sitter spacetime, 
with black hole and cosmological Killing horizons at radii $r_b$ and $r_c$ respectively.
Letting the subscript $h$ denote either horizon, the surface gravities $\kappa_h$ 
at the two horizons are found from the formula
$\kappa _h =  f_0^\prime (r_h) /2 $ to be
\be\label{kappah}
\kappa _b = \frac{\Lambda }{6 r_b} (r_c - r_b ) ( 2r_b + r_c ) \ ,  
\quad \kappa _c = - \frac{\Lambda }{6 r_c} (r_c - r_b ) ( 2r_c + r_b )
\ee
where we note that $\kappa_c$ is negative. The horizon temperatures are 
given by $T_h = |\kappa _h | /2\pi$ and the horizon entropies
are given by $GS_h = A_h/4 = \pi r_h ^2$.   

The thermodynamic volume $V$ is another relevant thermodynamic quantity,  
arising as the coefficient of the $\delta \Lambda$ term when the first law of 
black hole thermodynamics is extended to include variations in the cosmological 
constant \cite{Kastor:2009wy}. See \cite{Kubiznak:2016qmn} for an excellent 
review on this topic. As shown in \cite{Dolan:2013ft}, two different 
first laws may be proved for de Sitter black holes.  The derivation of the first, 
and less familiar, of these focuses on the region between the black hole and 
de Sitter horizons, and relates the variations in areas of the two horizons to 
the variation of the cosmological constant:
\be\label{firstsds}
T_b\,\delta S_b +T_c\,\delta S_c + M_p^2 V_{dS}\delta\Lambda = 0\,.
\ee
We refer to this statement as the \emph{SdS-patch first law}.  It holds for 
arbitrary perturbations around SdS that satisfy the vacuum Einstein 
equations with cosmological constant $\Lambda$.
The thermodynamic volume $V_{dS}$ for an SdS black hole works 
out to be simply the Euclidean volume between the horizons, which 
can be written equivalently in the two forms
\be\label{thermv}
V_{dS}= \frac{4\pi }{3} \left( r_c^3 - r_b^3 \right) \  
= \frac{4\pi }{\Lambda } \left( r_c - r_b \right)
\ee
In particular, even though the de Sitter patch first law \eqref{firstsds} does 
not include an explicit $\delta M$ term, it holds for perturbations within the 
SdS family in which both the parameters $M$ and $\Lambda$, or equivalently 
$r_b$ and $r_c$, are varied,  as can be verified straightforwardly using the 
formulae above. An alternative approach to a de Sitter patch first law is 
contained in \cite{Urano:2009xn}, in which the volume is varied.
An SdS patch Smarr formula \cite{Smarr:1972kt,Dolan:2013ft}, 
can be obtained by integrating the first law \eqref{firstsds} with respect to 
scale transformations, giving (in a general dimension $D$) 
\be
(D-2) T_bS_b +(D-2) T_cS_c -2M_p^2 V_{dS}\Lambda = 0
\ee
where the factors of $(D-2)$ and $-2$ arise respectively from the scaling 
dimensions of the horizon entropies and cosmological constant. In this paper 
we will be working in $D=4$.
(See also \cite{Sekiwa:2006qj} and references therein
for earlier ``phenomenological" arguments for an SdS Smarr relation.)

A second, independent first law, which we call the \emph{mass first law}, can 
be derived \cite{Dolan:2013ft} by considering a spatial region that stretches 
from the black hole horizon out to spatial infinity, and is given by
\be\label{firstmasssds}
\delta M-T_b\delta S_b + M_p^2V_b\delta\Lambda =0
\ee
The thermodynamic volume $V_b$ that arises in the mass first law is given in SdS 
by the Euclidean volume of the black hole horizon\footnote{A third first law, which 
we call the \emph{cosmological first law}, can also be obtained by considering the 
spatial region stretching from the cosmological horizon out to spatial infinity. 
The cosmological first law has the same form as \eqref{firstmasssds} with the 
subscript $b$ replaced by $c$, and with the corresponding thermodynamic 
volume given in SdS by $V_c =\frac{4\pi }{3}  r_c^3$. However, only two of 
the first laws are independent.  For example, subtracting the black hole and 
cosmological horizon versions, the mass term drops out giving the relation 
\eqref{firstsds} that only involves the geometry of the horizons.}, 
$V_b =\frac{4\pi }{3}  r_b^3$. Integrating the mass first law \eqref{firstmasssds} 
also gives a second independent Smarr formula
\be
(D-3)M-(D-2)T_bS_b -2 M_p^2 V_b\Lambda =0
\ee

\subsection{Convenient time coordinate for slow roll dynamics}
\label{coords}

As in \cite{Gregory:2017sor}, we now suppose that $\Lambda$ is an 
effective cosmological constant that varies in time as a scalar field, 
described by the action \eqref{action}, evolves in its potential $W(\phi )$. 
Our first step is to identify a preferred time coordinate $T$, such that 
the value of the scalar field throughout the spacetime depends only 
on $T$, and hence makes viable a scenario in which the effective 
cosmological constant $\Lambda$ also only depends on this time.
Such a time coordinate was identified for perturbative slow-roll evolution 
with an exponential potential in \cite{Chadburn:2013mta}, and for a general
slow transition in \cite{Gregory:2017sor}, 
\begin{equation}\label{xrs}
x= t - \frac{1}{2\kappa _c } \ln \left ( \frac{r_c -r}{r_c} \right ) 
+ \frac{1}{2\kappa _b } \ln \left ( \frac{r -r_b}{r_b } \right) 
+ \left ( \frac{r_c}{ 2\kappa_b r_b} -\frac{1}{2\kappa_n} \right ) 
\ln \left | \frac{r -r_n}{ r_n } \right | + \frac{r_b r_c }{ r_c -r_b }  \ln\frac{r}{r_0 } 
\end{equation}
where $r_n =- (r_b +r_c )$, as above, is the negative root of the SdS 
metric function$f_0(r)$, $\kappa_n = f_0^\prime (r_n)/2$ is the surface 
gravity evaluated at this root, and $r_0$ is an arbitrary constant of integration. 

\begin{figure}
\includegraphics[scale=0.8]{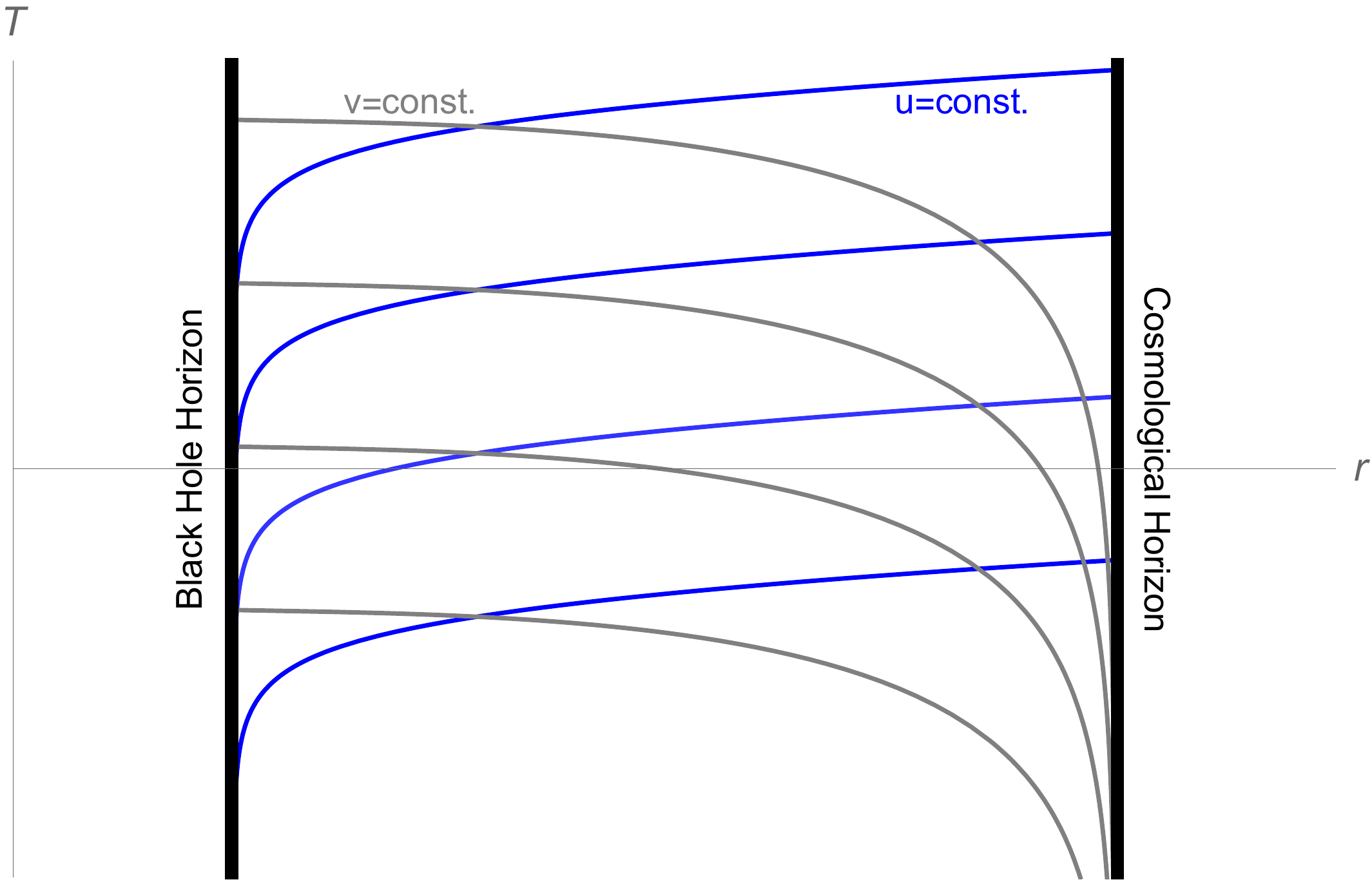}
\caption{A plot showing how the null lines $u=t-r^\star=$const,
$v=t+r^\star=$const appear in the $(T,r)$ coordinate system. 
Note how the $v$-lines meet the black hole horizon and the $u$-lines
the cosmological horizon at finite $T$.
}
\label{fig:uvlines}
\end{figure}

Starting from first principles  we review the argument that leads to our 
choice of a new time coordinate, which then 
gives SdS in an interesting new set of coordinates.
Let us begin by making a coordinate transformation of the SdS metric 
\eqref{sds} 
\begin{equation}\label{newcoord}
T = t + h(r)\  , \quad  
\end{equation}
where, for now, $h(r)$ is an arbitrary function of radius.
The SdS metric then has the form\footnote{Note that for the Schwarzschild 
metric, the choice $\frac{d h}{d r }f =\pm 1$ corresponds to outgoing/ingoing 
Eddington-Finkelstein coordinates.}
\begin{equation}\label{sdstwo}
ds^2 = -\fsds\, dT ^2 +  2\fsds h^\prime\, dT dr + \frac{1}{\fsds }
\left( 1-\left(\fsds h^\prime\right)^2 \right)dr^2 +r^2 d\Omega^2
\end{equation}
We look for solutions to the scalar wave equation \eqref{scalar} in these 
coordinates with $\phi=\phi(T)$, so that the wave equation then reduces to
\begin{equation}\label{wave} 
- \frac{1}{f_0} (1-\left( f_0 h^\prime\right)^2 ) \ddot{\phi} 
+ \frac{1}{r^2}\partial _r ( r^2 f_0  h^\prime ) \dot{\phi }
= \frac{d W }{d \phi}
\end{equation}
Assuming that the scalar field evolution may be approximated as a slow roll, 
we neglect the $\ddot{\phi }$ term on the left hand side. 
Under the assumption that $\phi=\phi(T)$, the right hand side of the wave 
equation depends only on $T$ and hence it is necessary that
\begin{equation}\label{condition}
\frac{1}{r^2}\partial _r ( r^2 f h^\prime ) = -3 \gamma 
\end{equation}
where $\gamma$ is a constant to be determined, and the factor of $-3$ 
is included for convenience. The wave equation then becomes
\be\label{wavetwo}
-3\gamma   \dot{\phi}  = \frac{d W}{ d\phi}\;,
\ee
giving $3\gamma$ the interpretation of a damping coefficient.
Equation \eqref{condition} can be integrated to obtain the required function 
$h(r)$ in the coordinate transformation \eqref{newcoord}, with the result
\begin{equation}\label{gett}
h^\prime = \frac{1}{f_0} \left( -\gamma r + \frac{\beta}{r^2}\right)  
\end{equation}
Here, $\beta$ is a further integration constant that is determined,
along with $\gamma$, by regularity at the two horizons\footnote{The 
constants $\beta$ and $\gamma$ will be key features of the slowly 
evolving black hole solutions found in Section \ref{solutionsection}, 
determining the relation between the rates of change of the mass 
and cosmological constant.}.
Specifically, we require that 
a solution $\phi(T)$ be an ingoing wave at the black hole horizon 
and an outgoing wave at the cosmological horizon, so that the 
scalar field is regular and physical in the SdS bulk. These conditions 
can be understood in terms of Kruskal-type coordinates which are 
smooth at the horizons 
\begin{equation}\label{kruskal}
U_h = - \frac{1}{|\kappa_h|} e^{-|\kappa_h | (t-r^\star )}  \  \  , 
\quad V_h = \frac{1}{ \kappa_h}  e^{\kappa_h (t+r^\star )}
\end{equation}
where the subscript $h=(b,c)$ designates either the black hole ($b$) 
or cosmological ($c$) horizons, and $r^\star$ is the tortoise coordinate
defined by 
\be\label{tort}
dr^\star = \frac{dr}{f_0(r)}
\ee
The black hole horizon is located at  $r^\star = -\infty$, while the 
cosmological horizon is at $r^\star =+\infty$.
It follows that $\ub =0$ on the black hole horizon, with $\vb$ 
a coordinate along the horizon.  
For the cosmological horizon, the situation is reversed with $\vc =0$ on 
the horizon and $\uc$ a coordinate along the horizon. Hence a regular solution
for the scalar field must behave like $\phi(T)\simeq \phi(t+r^\star)$
near the black hole horizon, and $\phi(T)\simeq \phi(t-r^\star)$
near the cosmological horizon. This implies that the function $h(r)$ 
in the coordinate transformation \eqref{newcoord} has the boundary 
conditions $h(r)\simeq +r^\star$ and $h(r) \simeq -r^\star$ as $r$ approaches 
$r_b$ and $r_c$ respectively.  It follows from the definition of the 
tortoise coordinate \eqref{tort} that the derivative of $h$ must behave like
\be\label{limits}
h^\prime (r) \simeq \begin{cases}
\frac{+1}{2\kappa_b(r-r_b)},\qquad r\rightarrow r_b\\
\frac{-1}{2\kappa_c(r-r_c)},\qquad r\rightarrow r_c
\end{cases}
\ee
On the other hand, equation \eqref{gett} implies that as  $r$ approaches 
$r_b$ or $r_c$, we must have
\be\label{limits}
h^\prime (r) \simeq \begin{cases}
\frac{+1}{2\kappa_b(r-r_b)} \left( -\gamma r_b +\frac{\beta}{r_b^2}\right ),
\qquad r\rightarrow r_b\\
\frac{+1}{2\kappa_c(r-r_c)}\left ( -\gamma r_c +\frac{\beta}{r_c^2} \right),
\qquad r\rightarrow r_c
\end{cases}
\ee
Comparing these two conditions we see that the constants 
$\beta$ and $\gamma$ must satisfy
\be\label{betagamma}
-\gamma r_b + \frac{\beta}{r_b^2}=1,\qquad 
-\gamma r_c + \frac{\beta}{r_c^2}=-1
\ee
which we can solve to obtain
\begin{equation}\label{alphabeta}
\gamma = \frac{ r_c ^2 + r_b ^2}{r_c ^3 - r_b ^3}\  \   , 
\quad \beta = \frac{r_c^2 r_b^2 (r_c  + r_b )}{r_c ^3 - r_b ^3 }
\end{equation}
Having found these expressions for $\gamma$ and $\beta$, we can now 
integrate equation \eqref{gett} and determine the function $h(r)$ that 
specifies the time coordinate $T$.  After some algebra, one finds that $T$
agrees with the coordinate $x$ in \eqref{xrs}, discovered in the specific case 
of an exponential scalar field potential \cite{Chadburn:2013mta}.
Combining the formulae \eqref{sdstwo} and \eqref{gett} now gives the SdS 
metric in the \emph{stationary patch} coordinates $(T, r)$,
\begin{equation}\label{sdsthree}
ds^2 = -\fsds dT^2 +  2\eta_{0} dr dT 
+ \frac{dr^2 }{\fsds} (1-\eta_{0}^2) +r^2 d\Omega^2
\end{equation}
where
\begin{equation}\label{etasdsdef}
\fsds h^\prime\equiv \eta_{0} (r) =   -\gamma  r +\frac{\beta }{r^2}
\end{equation}
with $\gamma$ and $\beta$ as given in \eqref{alphabeta}. 
Noting that $\eta_{0} (r_b ) =1$ and $\eta_{0} (r_c )= -1$, one sees 
that the boundary conditions on the scalar field have yielded the result  
that near the black hole horizon, the time coordinate $T$ coincides with 
the ingoing Eddington-Finkelstein coordinate, while near the cosmological 
horizon it becomes the outgoing coordinate. Hence the $(T,r)$ coordinates 
are well behaved in the region between and on the horizons.  
Figure \ref{fig:uvlines} shows schematically how
the null $(t\pm r^\star)$ coordinates $(u,v)$ appear in this alternate
system. These properties will be useful when analyzing the near-horizon 
behavior of our dynamical solutions in Section \ref{nhb}. 

The constants $\gamma$ and $\beta$ are both positive, 
and the damping coefficient $3\gamma$ in the evolution equation 
\eqref{wavetwo} for the scalar field has a simple geometrical interpretation 
as the ratio of the sum of the black hole and cosmological horizon areas 
$A_{total} = 4\pi ( r_c ^2 + r_b ^2 )$  to the thermodynamic volume 
$V_{dS}= 4\pi ( r_c ^3 - r_b ^3 )/3$ from the SdS patch first law \eqref{firstsds},
\be\label{gammainterp}
3\gamma = \frac{A_{total}}{V_{dS}}\,.
\ee
as was noted in \cite{Gregory:2017sor}. The damping of the scalar 
field evolution is greatest for small $V_{dS}$, 
which corresponds to a large black hole with radius approaching that 
of the cosmological horizon.

\subsection{Interlude on slow-roll approximation}\label{slowroll}

Before proceeding with finding the solution, we outline more precisely 
the approximations within which we will be working. The resulting 
conditions on the potential are essentially the same as those in slow-roll inflation 
without a black hole \cite{Liddle:1993fq,Liddle:1994dx}, summarized 
in equations \eqref{sloweps} and \eqref{slowdelta} below.

The picture is that we have a scalar potential with a shallow gradient 
so that the kinetic contribution of the scalar energy momentum tensor 
and the time-dependent corrections to $W$ are sub-dominant to the
zeroth order contribution from the potential. In inflation, we summarise 
the slow-roll conditions in terms of the acceleration of the universe: 
``$\dot{H}/H<1$''. Here however we are in stationary coordinates for 
SdS spacetime, therefore we must identify our slow-roll in a slightly 
different way. However, the useful parameters turn out to be the same 
as the conditions familiar for typical inflationary models.

Similar to slow roll inflation, we will demand that the stress-energy is 
dominated by the contribution of the scalar potential, and that the 
$\ddot{\phi}$ term be subdominant to the $\dot{\phi}$ term
in the scalar equation of motion. That is,
\be
\frac{1-\eta^2}{f} {\dot{\phi}}^2 \ll W \;\;,\qquad
\frac{1-\eta^2}{f} {\ddot{\phi}} \ll \frac1{r^2}\left |  \left ( r^2 \eta\right)' \dot{\phi}\right|
\ee
Since ${\dot{\phi}}$ is assumed perturbatively small, the metric coefficients in the 
above relations can be taken to be their zeroth order values. Note that
$(1-\eta^2_0)/f_0$ is monotonically decreasing between the two horizons,
and $(r^2\eta_0)'/r^2 = -3\gamma$ is constant. Hence we can manipulate the first
relation to give
\be
\frac{1-\eta_0^2}{f_0} \frac{{\dot{\phi}}^2}{W} 
< -\frac{\eta '(r_b)}{\kappa_b}  \frac{{\dot{\phi}}^2}{W} 
= \frac{2}{3\Lambda} \frac{(2r_c^3+3r_c^2r_b+r_b^3)(r_b^2+r_c^2+r_cr_b)}
{(2r_b+r_c)(r_b^2+r_c^2)^2} \frac{W^{\prime2}}{W} 
\simeq M_p^2 \frac{W^{\prime2}}{W^2}\ll1
\ee
suggesting that we use the slow-roll parameter
\be\label{sloweps}
\varepsilon = M_p^2 \frac{W^{\prime2}}{W^2}\ll1
\ee
to quantify how ``small'' the kinetic term of the scalar is: 
$\dps \lesssim \varepsilon W$.

For the second relation, we take the leading order approximation to
the $\phi-$equation, \eqref{wavetwo}, and differentiate to find $\ddot{\phi}$, 
giving
\be
\left | \frac{1-\eta_0^2}{3\gamma f_0} \frac{\ddot{\phi}}{\dot{\phi}} \right | <
\frac{2}{3\Lambda} \frac{(2r_c^3+3r_c^2r_b+r_b^3)(r_b^2+r_c^2+r_cr_b)}
{(2r_b+r_c)(r_b^2+r_c^2)^2} W'' \simeq M_p^2 \frac{W''}{W} \ll 1
\ee
suggesting again that we use the parameter
\be\label{slowdelta}
\delta = M_p^2 \frac{W''}{W}\ll1
\ee
to quantify how small the variation of the kinetic term is:
$\ddot{\phi} \lesssim \dot{\phi} \delta/r_c$. Note that this
is again similar to the `eta' parameter of the traditional slow-roll approach,
here renamed as $\delta$. 

We see therefore that the slow-roll relations are more or less the same
as for the FRW cosmology without a black hole, however, the black 
hole introduces a radial dependence in the stationary patch coordinates
that is easily bounded.
The main difference between slow-roll with and without a 
black hole is the $\gamma-$parameter responsible for the friction of
the scalar field. Since $\Lambda = {\cal O}(r_c^{-2})$, 
$\gamma = {\cal O}(\sqrt{\Lambda})/(1-r_b/r_c)$ can be
very much greater than the Hubble parameter that usually serves
as the friction term in inflationary slow-roll, although it should be noted
that the time with respect to which the scalar is rolling is the stationary 
patch time coordinate $T$.

\section{Solution for the Dynamical Metric}
\label{solutionsection}

In the last section we analyzed the slow-roll evolution of a test 
scalar field $\phi$ with potential $W(\phi)$ in a background SdS 
spacetime. We found a suitable time coordinate $T$, such that 
there exist solutions $\phi=\phi(T)$ that are purely ingoing at the 
black hole horizon and outgoing at the cosmological horizon. 
Our next goal is to solve the back-reaction problem in perturbation 
theory to obtain the leading corrections to the SdS black hole metric 
as the scalar field rolls down its potential.  We will present a preview 
of the results here and then provide their derivation in the next subsection.
We start by making an ansatz for the form of the metric, taking the 
form of the SdS metric in the $(T,r)$ coordinates \eqref{sdsthree} but 
allowing the metric functions to depend on both the $T$ and $r$ coordinates
\begin{equation}\label{metric}
ds^2 = -f dT^2 +  2\eta dr dT + \frac{dr^2 }{f} ( 1-\eta^2  ) +r^2 d\Omega^2
\end{equation}
where $f=f(r,T)$ and $\eta = \eta (r,t)$. To be specific, we
assume that the scalar field starts at time $T=T_0$ with a value 
$\phi_0$ corresponding to a maximum of the potential.  The initial 
cosmological constant is given by $\Lambda_0 = W(\phi_0)/M_p^2$,
and we assume that we start with an SdS black hole, so that the 
unperturbed metric has the form \eqref{sdsthree} with
\be 
f(r,T) = f_0  (r; M_0, \Lambda_0 )= 1- \frac{2GM_0}{ r}- \frac{\Lambda_0}{3}r^2
\ee
where $M_0$ is the initial black hole mass. 
Corresponding to the values of $M_0$, $\Lambda_0$ are initial 
values $r_{b0}$, $r_{c0 }$ of the black hole and cosmological horizon 
radii, which in turn give the parameters $\beta$, $\gamma$ for the 
unperturbed SdS spacetime via \eqref{alphabeta}.

Moving forward in time, the scalar field $\phi$ evolves via the scalar 
equation. In the background SdS spacetime, $\phi$ is constant, and 
therefore the time-dependent part of $\phi$ is perturbative and to 
leading order evolves according to the wave equation \eqref{wavetwo} 
in the background SdS spacetime. We note for future use that the time 
coordinate is related to the evolving scalar field by
\be\label{slowphi}
T= - \frac{1}{3\gamma}\int \frac{d\phi}{(d W / d\phi )}
\ee

With this set-up, we can now state our results for the evolution of the 
spacetime geometry. The physical picture is that for slow-roll evolutions, 
the metric should transition through a progression of SdS metrics, with 
the parameters $M$ and $\Lambda$ varying slowly in time. We will show 
that the solutions come very close to realizing this expectation. Explicitly, we 
find that to leading order in perturbation theory, a metric of the form 
\eqref{metric} satisfies the Einstein equations with stress-energy coming 
from the rolling scalar field, and metric functions given by
\be\label{perturbed}
f =  f_{QS} (r,T)  + \delta f (r,T) \  ,\qquad 
\eta =  \eta_{0}(r) + \delta \eta (T,r) 
\ee
where
\be
\beal
f_{QS} (r,T)  & =  f_0  \left ( r; M(T) , \Lambda (T)  \right )
=  1  - \frac{2GM(T)}{r} - \frac{\Lambda (T)}{3}  r^2 \\
\delta f(r,T) &= - \frac{1}{2r} \frac{\dot{\phi}^2}{M_p^2} \int_{r_{b0}} ^r   
\frac{ 1-\eta_0^2}{f_0 }r'^2dr',\qquad   
\delta \eta(r,T) = \frac{r ( 1-\eta_0^2 )}{2 f_0 } 
\int_{T_0}^T  \frac{\dot{\phi}^2}{M_p^2} dT
\eeal
\label{deltas}
\ee
with
\be
\dot{M}  = \frac{dM}{dT}
= 4\pi \beta \dot{\phi}^2 ,  \qquad 
\dot\Lambda = -3\gamma \frac{\dot{\phi}^2}{M_p^2}
\label{derivs}
\ee
We will also demonstrate that the function $\delta f$ above is transient,
and sub-dominant to the time dependent part of $f_{QS} (r, T)$.
Thus the time dependent black hole and cosmological horizon radii $r_h (T)$
are well approximated by the zeros $r_h^{QS} (T)$ of the quasi-statically 
evolving portion $f_{QS} (r, T)$. By construction these zeros are
related to the time dependent mass and cosmological constant $M(T)$ 
and $\Lambda (T)$ according to the SdS formulae in \eqref{mlambda}. 
Hence the time derivatives of $M(T) $ and $\Lambda (T)$ in \eqref{derivs} 
can be converted into time derivatives of the horizon radii $r_h (T)$, giving
\be\label{rhdot} 
\dot{r}_h = \frac{ r_h}{2 |\kappa _h | } \frac{\dps}{M_p^2}  \  , \quad h=(b,c)
\ee
Expressions for the time dependent mass and cosmological constant can 
be obtained by using the integral
\be\label{phiintegral}
\int_{T_0}^T \dot{\phi}^2 dT' =
-{1\over 3\gamma}\left[ W(\phi(T)) - W(\phi (T_0))\right]
\ee
Integrating $\dot M$, $\dot\Lambda$ and $\dot r_h$  using 
\eqref{phiintegral} we arrive at the results 
\be\label{solextra}
\Lambda (T)\approx W(\phi(T) ) \   , \qquad  M(T)    
\approx M_0+{\beta\over 6\gamma} \left[ W_0-W(\phi (T )) \right] 
\ee
and
\be\label{solextratwo}
r_h \approx r_{h0} +  \frac{ r_{h0} }{6\gamma |\kappa _h | } 
\left[ W_0- W(\phi (T ))\right] 
\ee
and similarly for the function $\delta\eta(r,T)$ in \eqref{deltas}. 
Note, we use the approximate equality above, as in the slow
roll approximation it is possible for $M$ and $\Lambda$ to
have additional slowly varying contributions (such as $\dps$) whose derivatives are order $\ddot{\phi } $ or higher, and hence
would not contribute to \eqref{derivs} at this order. We have 
denoted here $T_0$ as an initial time, which could be at 
$T_0 \rightarrow -\infty$, when the scalar field is at the top of the 
potential with value $\phi_0$, the metric is SdS with mass parameter 
$M_0$, $\Lambda_0 = W(\phi_0 )/M_p^2$, 
and the horizon radii are the corresponding  $r_{h0}$  determined by 
\eqref{mlambda}. Note that the value of the scalar field potential, and 
hence the effective cosmological constant, is decreasing in time, so 
that the mass and both horizon radii are increasing functions of time.

\subsection{Solving the Einstein equations}
\label{findsolution}

We now present the derivation of the quasi-de Sitter black hole solutions 
given above. With the assumption that
$\phi = \phi(T)$, there are two independent components of the stress tensor,
\be
\beal
T_{TT}  &= \left ( W (\phi ) + \frac{ 1+\eta^2}{2f }\dot{\phi}^2 \right) |g_{TT}|
\;, \qquad \text{and} \\ 
T_{\alpha\beta} &= \left(-W(\phi ) +  \frac{ 1-\eta^2 }{2f }\dot{\phi}^2\right)  
g_{\alpha\beta} \qquad \text{otherwise.}
\eeal
\label{tabortho}
\ee
and the Einstein tensor is:
\be
\beal
G_{TT} & = \left [ \frac{1}{r^2} ( 1-f-rf^\prime ) -  \frac{\eta \df }{rf} 
\right] |g_{TT}| \\ 
G_{rr} & = \left [ -\frac{1}{r^2} ( 1-f-rf^\prime ) +  \frac{\eta \df }{rf}  
+ \frac{2\dot{\eta}}{r(1-\eta ^2 )}  \right] g_{rr}  \\  
G_{rT} & =  \left [ -\frac{1}{r^2} ( 1-f-rf^\prime )  
-  \frac{ (1-\eta^2 ) \df}{r\eta f} \right] g_{rT} \\
G_{\theta\theta}  & = \left [ \frac{f^{\prime\prime}}{2} + \frac{f'}{r} 
-\frac{\eta^\prime\df}{2f}  + \frac{\dot{\eta} f^\prime}{2f} + \frac{\dot{\eta}}{r}
+ \dot{\eta}^\prime+ \frac12 \left ( \frac{(\eta^2-1)}{f} 
\right)^{\!\boldsymbol{\cdot}\boldsymbol{\cdot}}
\right] g_{\theta\theta}
=  \frac{G_{\phi\phi}}{\sin^2\theta} 
\eeal
\label{einstein}
\ee

First consider the linear combination $\frac{G_{TT}}{|g_{TT}|} +\frac{G_{Tr}}{g_{Tr}} 
= M_p^{-2} \left (\frac{T_{TT }}{|g_{TT}|} + \frac{T_{Tr}}{g_{rT}} \right)$, 
which gives the relation
\be\label{fdot}
\df =-r\eta\, \frac{\dps}{M_p^2}
\ee
and then $\frac{G_{TT}}{|g_{TT}|} +\frac{G_{rr}}{g_{rr}} 
= M_p^{-2} \left (\frac{T_{TT }}{|g_{TT}|} + \frac{T_{rr}}{g_{rr}} \right)$,
which yields 
\be\label{etadot}
\deta = \frac{ r(1-\eta^2 )}{2f}\,\frac{\dps}{M_p^2}
\ee
thus both of our metric functions have a time-dependence determined
by a radial profile times the kinetic energy of the scalar field. 
Since the quantity $\dot\phi^2$ is already perturbative, we need only
consider the metric functions on the right hand sides of equations 
\eqref{fdot} and \eqref{etadot} to leading order, i.e.\ $f\to f_0$ and 
$\eta\to\eta_0$, the zeroth order SdS functions given in \eqref{sds} 
and \eqref{etasdsdef}. So in perturbation theory, these 
equations give explicit expressions for the time evolution of  $f$ and $\eta$.

Equations \eqref{fdot} and \eqref{etadot} can now be substituted back 
into the Einstein tensor in \eqref{einstein} to eliminate $\df$ and $\deta$.
One then finds that the $TT$, $T r$, and $rr$ components of the 
normalized Einstein equation all reduce to
\be\label{oneeinstein}
\frac{1}{r^2} ( 1-f-rf^\prime ) = \frac{W(\phi)}{M_p^2} 
+ \frac{(1-\eta_0^2)}{2f_0 }\,  \frac{\dot{\phi}^2}{M_p^2} 
\ee
There remains the $\theta\theta$ equation, which using  \eqref{fdot} 
and \eqref{etadot} becomes
\be\label{thth}
- \frac12 f^{\prime\prime} - \frac{f^\prime}{r}=  \frac{W }{M_p^2}
+ \frac{1}{2f_0}  \left[ (1-\eta_0^2 ) -r\eta_0\eta_0^\prime 
-r(1-\eta_0^2 ) f_0^\prime \right]\, \frac{\dot{\phi}^2}{M_p^2}
\ee
Taking the derivative with respect to $r$ of the Einstein equation 
\eqref{oneeinstein} gives \eqref{thth}. So the equations of motion for 
the coupled Einstein scalar field system have been reduced
to four equations, namely \eqref{fdot}, \eqref{etadot} and 
\eqref{oneeinstein}, together with the slow-roll equation \eqref{wavetwo} 
for the scalar field.

Physical considerations now give insight as to the 
perturbative form of the metric functions $f(r,T)$ and $\eta(r,T)$.
During slow roll evolution in an inflationary universe without a black hole
the scalar field potential $W(\phi )$ provides a slowly varying effective 
cosmological constant. The stress-energy is 
dominated by the potential, with small corrections coming from the 
kinetic contribution of the scalar field.
As the potential changes, the metric is approximately given by a 
progression of de Sitter metrics.
With a black hole present, for sufficiently slow evolution, one expects 
that the metric will be close to an SdS metric with slowly evolving 
parameters $M(T)$ and $\Lambda(T)$. Looking at the expressions 
for the first three normalized components of the Einstein tensor \eqref{einstein},
one sees that if $f=f_0$, then the terms in the first round parentheses are all 
proportional to $\Lambda_0$. Since these terms only depend on 
radial derivatives, if we use the `quasi-static' metric function 
$f_{QS} (r, T)$ defined in \eqref{deltas}
then these terms will be proportional to $\Lambda(T)$.
Further, looking also at the normalized stress-energy components 
\eqref{tabortho}, one sees that the Einstein equations suggest a 
strategy of balancing $\Lambda (T)$ with $W[\phi (T)]$ and then 
equating the remaining terms. These are corrections to a pure 
cosmological equation of state from the kinetic energy of the scalar field.
 
Based on this reasoning, we adopt an ansatz of the form
\be
\beal
f(r,T ) &= f_{QS} (r,T) + \delta f (r,T)  \\  
\eta (r,T ) &= \eta_{0} ( r) + \delta\eta (r,T) 
\eeal
\label{fdef}
\ee
It may appear redundant to have the time dependence for $f$ 
explicit in both the function $f_{QS} (r,T)$, as well as $ \delta f (r,T)$.
However, it turns out this is a convenient way of packaging the time
dependence implied by the Einstein equations in a physically relevant manner.
One might also question the asymmetry in \eqref{fdef} between $f(r,T)$ and 
$\eta(r,T)$.  However, this is simply what turns out to work. 
The next step is to substitute \eqref{fdef} into the Einstein equations, 
show that the ansatz allows a solution, and solve for the four functions 
$M(T)$, $\Lambda(T)$, $\delta f(r,T)$, and $\delta \eta(r,T)$.

Starting with $f(r,T)$, we substitute our ansatz into equations 
\eqref{oneeinstein} and \eqref{fdot}, to get
\bea
\Lambda (T) - {1\over r^2} \frac{\partial~}{\partial r } \left( r \delta f  \right) 
&=& \frac{W\left [ \phi (T)\right]}{M_p^2} 
+ \frac{(1-\eta_0^2)}{2f _0}\, \frac{\dot{\phi}^2}{M_p^2}
\label{einstone}\\
\frac{\dot{\Lambda} r^2}{3} + \frac{2G\dot{M}}{r} - \dot{\delta f}
&=& \left ( -\gamma r^2 + \frac{\beta}{r} \right) \frac{\dot{\phi}^2}{M_p^2}
\label{delfconsis}
\eea
respectively.
According to our physical picture, we try for a solution such that the 
cosmological constant tracks the value of the scalar potential, 
$\Lambda (T) = W[\phi (T)]/M_p^2$. Taking the time derivative of this,
and using the slow-roll equation \eqref{wavetwo} for the scalar field, 
we get
\be\label{lamdotconsis}
\dot{\Lambda } (T) 
= -3\gamma \frac{\dps}{M_p^2}\;,
\ee
consistent with the ${\cal O}(r^2)$ terms in \eqref{delfconsis}.

Next, we integrate \eqref{einstone} (after cancelling the $\Lambda$
and $W$ terms) to obtain
\be\label{soltwo}
\delta f (r,T)
=  - \frac{1}{2r} \left( \int ^r_{r_{b0}} dr^\prime (r^\prime )^2 
\frac{ 1-\eta_0^2}{f_0 }\right) \frac{\dot{\phi}^2}{M_p^2} 
\ee
Here, we have chosen for convenience to fix the lower limit of the
integral to be the initial black hole horizon radius $r_{b0}$, so that 
$\delta f (r_{0b},T ) =0$; an alternate lower
limit would correspond to a shift in $\delta f\propto \dps/r$ that without
loss of generality can be absorbed in $M(T)$ -- recall that within
the slow roll approximation, the rate of change of $\dps$ is ignorable.
Since second derivatives of $\phi$ are being neglected in the slow roll 
approximation, equation \eqref{soltwo} implies that to this order
\be\label{psidotone}
\frac{d~ }{dT} \delta f \simeq 0
\ee
and hence we read off from \eqref{delfconsis}
\be\label{mlambdadot}
\dot M(T) = 4\pi \beta\,\dps 
\ee 
This rate of change of $M$ corresponds to the
accretion of scalar matter into the black hole. 
We have therefore demonstrated the quasi-static form of the solution for $f$,
subject to the slowly varying $\delta f$.

It is interesting to compare this expression for $\dot M$ to that obtained
for fluid accretion onto an asymptotically flat black hole by Babichev et al.\
\cite{Babichev:2005py,Babichev:2012sg}, 
\be\label{mdotsmall}
\dot{M}(V) = - 4\pi r^2 T_V^r \;,
\ee
where $V$ is the Eddington-Finkelstein advanced time coordinate for
the black hole, and we have used the more general expression from 
\cite{Babichev:2012sg} to facilitate comparison. Taking the cosmological 
constant tending to zero, that is, $r_b\ll r_c$, we find that our expression for 
$T$ reduces to the Eddington-Finkelstein coordinate $V$, and $\beta \approx r_b^2$.
Finally, using the expression for the energy momentum tensor, 
$\dot{\phi}^2 = - T_V^r$ giving precisely the expression \eqref{mdotsmall}
of Babichev et al.

Finally, we integrate \eqref{etadot} to obtain
\be\label{solthree}
\delta\eta (r,T) =   \frac{r ( 1-\eta_0^2 )}{2 f_0 } 
\int_{T_0}^T  \frac{\dot{\phi}^2}{M_p^2}  dT^\prime
+\eta_1(r,T)
\ee
where $\eta_1$ is a slowly varying function ($\dot\eta_1\approx0$) 
that will turn out to be proportional to $\dps$, and we require that 
$\delta \eta$ is zero at the start of the slow roll 
process $T_0$.

This completes the derivation of the slow-roll solution.
The evolution of all the functions, $M(T)$, $\Lambda(T)$, 
$\delta f(r,T)$, and $\delta\eta(r,T)$ in the quasi-SdS metric \eqref{fdef} 
are determined in terms of the unperturbed SdS metric and $\dot{\phi} (T)$, 
which is in turn determined through the slow roll equation \eqref{wavetwo} 
by the slope $ d W /d \phi $ of the scalar potential. 
While the quasi-SdS solution is not quite as simple as the static SdS 
spacetime, we will see that the quasi-SdS form allows us to study the 
thermodynamics of both the black hole and cosmological horizons in an intuitive way.

\section{Horizons}\label{horsection}

In this section we locate the horizons $r_h (T)$ of the dynamical metric.
In the background static SdS metric, $M_0$ and $\Lambda_0$ are 
related to the horizon radii $r_{b0}$ and $r_{c0}$ by the formulae 
\eqref{mlambda}. Since our solutions almost track a sequence of SdS 
metrics we expect that the black hole and cosmological horizons
are given by almost the same SdS formulae, with $M(T)$ and $\Lambda (T)$ 
replacing the constant parameters. However, ``almost" is the operative word, 
since  both $f$ and $\eta$ receive modifications. Nonetheless, 
our findings are that:

\noindent $\bullet$ After an initial time period, the $\delta f$ corrections 
become negligible;

\noindent $\bullet$ $\eta (T, r) $ still satisfies the needed boundary
conditions, and as a result,

\noindent $\bullet$ the simple quasi-static formulae for the horizon radii 
apply after that initial period.

The derivation proceeds in two steps. We first find the time dependent zeros 
of $f(T, r)$ and identify the ``start up" time interval.
Second, we look at the form of the metric near the zeros of $f$ 
and show that these are horizons. We close this section by comparing the 
results to those in our previous paper \cite{Gregory:2017sor}  
which used a different coordinate system for the evolving system.

\subsection{Finding zeros of $f(r,T)$}
\label{findingzeros}

Here we show that after an initial time period estimated below, 
the horizons $r_h (T)$ are given by the zeros of $f_{QS} $, that is, the 
$r_h (T)$ with $h=(b,c)$,  have the same algebraic relation to $M(T)$ 
and $\Lambda (T)$ as do the quantities in static SdS. 
In the unperturbed SdS spacetime, the black hole and cosmological 
Killing horizons are located at the zeros of $f_{0}(r)$, so we start by finding the zeros of  
$f(r, T)$, which we designate by $r_{z,h} (T)$. That is, 
\be\label{fzeros}
f \left( r_{z,h} (T),T \right) =0
\ee
where $f(r,T)$ is given in equation \eqref{perturbed} by
\be\label{fagain}
f(r,T) =f_{QS} (r, T) - \frac{1}{2r} \frac{\dps}{M_p^2} I(r)
\ee
with
\be\label{integral}
I (r ) = \int _{r_{b0}}^r \frac{(1-\eta_0^2)}{f_0}  r^{\prime 2}dr^\prime 
\ee
Denote the zeroes of the quasi-static metric function $f_{QS}(r,T)$, 
defined in \eqref{deltas}, as $r_h^{QS}(T)$, so that
\be\label{fqsz}
f_{QS} \left (r_h^{QS} (T) , T\right ) =0 
\ee
It follows that the quasi-static radii $r_h^{QS}(T)$ are given in terms 
of the time dependent mass and cosmological constant parameters 
$M(T )$ and $\Lambda (T)$ by the SdS relations \eqref{mlambda}.

Let us now compare the $r_{z,h} (T)$ to the $  r_h^{QS} (T)$.
Since by construction $I(r_{b0} ) = 0$, these two types of zeroes 
coincide at the black hole horizon\footnote{Recall that the factor 
$\dot\phi^2$ preceding $I(r)$ in \eqref{fagain} is perturbative.  
It is therefore sufficient to evaluate the integral at the unperturbed 
horizon radius.}, 
\be\label{samerb}
r_{z,b} (T) = r_b^{QS} (T)
\ee
On the other hand, when $r=r_{c0}$ in $I$, equation \eqref{fzeros} becomes
\be\label{fzrc}
- \frac{\Lambda (T)}{3} r_{z,c}^3 + r_{z,c} -2GM(T) 
- \frac12 \frac{\dps}{M_p^2} I( r_{c0} ) =0
\ee
which is not  the same as the condition \eqref{fqsz}. Hence, the 
radii $r_c^{QS} (T) $ and $ r_{z,c} (T) $ differ, with the difference arising 
from $I(r_{c0})$, which enters the polynomial equation \eqref{fzrc} as a 
contribution to the time dependence of the  mass term $M(T)$.  
Let us compare the magnitudes of the last two terms in \eqref{fzrc}. 
One might be concerned that $I(r_{c0})$ would diverge due to the 
factor of $1/f_0$ in the integrand in \eqref{integral}.  However, the 
zeros of $1-\eta_0^2$ cancel the zeros of $f_0$ at $r_{b0}$ and 
$r_{c0}$.  Explicitly,
\be\label{oneminus}
1-\eta_0 = \frac{(r-r_{b0} )}{(r_{c0}^3 - r_{b0} ^3 ) r^2 } P(r, r_{b0} , r_{c0} ) \  , 
\quad 1+\eta_0 = \frac{(r_{c0} -r )}{(r_{c0}^3 - r_{b0} ^3 ) r^2  }P(r, r_{c0} , r_{b0} )
\ee
where $P(r, d, e) =  r^2 (d^2 + e^2 ) +r e^2 (d+ e) +de^2(d+e) $.  
Therefore the integral may be rewritten as
\be\label{integraltwo}
I (r) = \frac{3}{\Lambda (r_{c0}^3 - r_{b0} ^3 )^2  }  
\int _{r_{b0}}^r dr^\prime 
\frac{ P(r^\prime , r_{b0} , r_{c0} )P(r^\prime , r_{c0} , r_{b0} ) }
{ r^\prime (r^\prime -r_N) } 
\ee
which is manifestly finite at $r=r_{c0}$. So the final term in \eqref{fzrc} 
is then proportional to the instantaneous value of $\dot\phi ^2$ times 
this finite, time-independent, quantity, while from \eqref{derivs} we 
know that the change in $M(T)$ from its initial value $M_0$ depends 
on the accumulated change of  $\dot\phi ^2$ integrated over time (plus
a possible slowly varying term). Therefore, after a start up time the 
contribution to $r_{z,c}$ coming from $ \dot\phi ^2I(r_{c0})$ will be 
small compared to the contribution from the evolution of $[M(T)- M_0]$. 
Explicit comparison of these terms shows that  the shift due to the 
$I(r_{c0})$ term is  negligible\footnote{For clarity, we have dropped
numerical constants in making the following estimates.} after
a time interval  $(T- T_0) \simeq r_{b0} \ln (r_{c0} / r_{b0})$ for small 
black holes with $r_{b0} \ll r_{c0}$, and after $(T-T_0) \simeq r_b$ for 
large black hole with $r_{b0}$ of order $r_{c0}$.  Note that if the 
initial time $T_0$ is taken to be $ -\infty$ then the extra term is 
irrelevant\footnote{Naturally, a different gauge choice could have been 
made to set $I =0$ at $r_c$, and the correction would show up at $r_b$.}.
After this initial time period the zeros of $f(r,T)$ coincide with the 
zeros of $f_{QS}(r,T)$ 
\be\label{latehorizons}
r_{z,b}(T) = r_b^{QS} (T)  \  , \quad r_{z,c}(T) = r_c ^{QS}(T)
\ee
and hence evolve in a quasi-static way determined by the evolution 
of the quasi-SdS parameters $M(T)$ and $\Lambda(T)$. It may seem 
odd that the additional function $\delta f$ is needed to solve the Einstein
equation at early times, rather than at late times. This likely illustrates 
the teleological nature of the horizon, as it starts to grow in anticipation 
of the influx of stress-energy from the evolving scalar field. This effect 
was also noticed in a simpler inflationary example with no black hole
in \cite{Gregory:2017sor}.
  
\subsection{Near horizon behavior}\label{nhb}

In this section we find the form of the metric near the zeros of $f$ and 
show that it takes the form of a black hole or cosmological horizon. 
This then implies that the functions $ r_{z,h}(T)$ do locate the horizons, 
and combining with the results of the previous section, we will have 
demonstrated that  the horizon radii evolve quasi-statically,
\be\label{allthesame}
r_h (T) = r_{z,h}(T) \simeq r_h^{QS}(T)
\ee
after an initial period as discussed above.

Let us recall what the horizons look like in the static SdS metric, 
starting with the static time coordinate $t$, and $r$.
Defining the  radial tortoise coordinate by $dr^\star = dr /f_{0}$ and 
the ingoing Eddington-Finklestein coordinate at the black hole horizon 
by $v= t+ r^\star$, the SdS metric  near the black hole horizon is 
approximately given by
\be\label{sdsbh}
ds^2 \simeq -2\kappa_b (r- r_b ) dv^2 + 2drdv +r^2d\Omega^2
\ee
Although the metric component $g_{vv}\simeq -2\kappa_b (r- r_b ) dv^2$ 
vanishes at the horizon, the metric is non-degenerate there. The vector 
$\partial / \partial v$ is null and ingoing at the horizon. To study the 
cosmological horizon, one uses outgoing coordinates with $u=t-r^\star$. 
Then near the cosmological horizon the metric becomes
\be\label{sdsch}
ds^2 \simeq 2|\kappa_c | (r- r_c ) dv^2 - 2drdu +r^2d\Omega^2
\ee
The vector field $\partial / \partial u$ is outgoing and  null at the horizon and the cross term in the metric changes sign.

One can also look at the metric near the horizons of SdS using the
the stationary coordinate $T$ given in \eqref{xrs}. As noted earlier, 
$T$ interpolates between the ingoing null Eddington-Finklestein 
coordinate $v$ on the black hole horizon, and the outgoing
null coordinate $u$ on the cosmological horizon. So there is no 
need to transform the coordinates, we just look at the metric functions 
near the horizons. As $r\rightarrow r_b$ the SdS metric becomes
\be\label{hormetric}
ds^2 \simeq - 2\kappa_b (r- r_b ) dT^2 + 2dTdr 
+ \frac{C_b}{\kappa _b } dr^2 +r^2 d\Omega^2 \  
\ee
where $C_b = -\eta^\prime_{0} (r_b ) = ( 2r_c^3 +3 r_c^2 r_b + r_b^3 )
[r_b (r_c^3 - r_b^3 )]^{-1} \ >0$.
The essential difference between \eqref{sdsbh} and \eqref{hormetric} 
is that $g_{rr} =0$ in the $(v,r)$ coordinates, while $g_{rr} $ is non-zero 
but finite in the $(T,r)$ coordinates. This is due to the fact that  $T$ is 
not identical to $v$ near $r_b$. Similarly, near the cosmological horizon 
of SdS the metric in stationary cooordinates has the form
\be\label{coshormetric}
ds^2 \simeq  2| \kappa_c | (r- r_c ) dT^2 - 2dTdr 
+ \frac{ C_c }{ | \kappa _c | } dr^2 +r^2 d\Omega^2 \  
\ee
and one sees explicitly that $T$ has become an outgoing null coordinate. 
Here $C_c = -\eta^\prime_{0} (r_c )>0$.
 
The analysis of the time-dependent metric in \eqref{perturbed}  near the 
zeros of $f$ proceeds in a similar way.
Take the time $T$ to be sufficiently large that the zeros of $f$ are
well approximated by the zeros of $f_{QS}$. Further,
to avoid repetition of bulky notation, we will simply abbreviate 
$r_h ^{QS} (T) = r_h (T)$, and the end result
justifies this replacement. 
As $r$ approaches $r_b (T)$, the metric function $f$ behaves like
\be\label{nearhf}
f\simeq - 2\kappa_b (T)  \left(r-r_b (T) \right) 
\ee
where $\kappa_b (T)$ is defined as the derivative
\be\label{kbdef}
\kappa_b (T) = {1\over 2} f^\prime |_{r=r_b(T)} 
\ee
At present we will not ascribe any significance to $\kappa_b (T)$ as a 
possible time-dependent surface gravity or temperature. 

The behavior of the metric function
\be\label{grr}
g_{rr} = \frac{ (1-\eta^2 )}{f}
\ee
in the time dependent solution \eqref{perturbed} appears to be more subtle, 
since although the zeros of $f$ in the denominator are cancelled by zeros 
in the numerator for exact SdS, it is not immediately obvious whether 
this is still true for  the perturbed functions $f$ and $\eta$. However, 
\eqref{fdot} and \eqref{etadot} together imply that $g_{rr}$ is in fact
independent of $T$, provided only that $\phi = \phi(T)$, hence $g_{rr}$
remains regular even if the locations of the zeros of $f$ shift.
Let us now confirm this, by deriving explicitly the locations of these
zeros. Denote $r_{h0}$ as an horizon radius in the 
unperturbed SdS spacetime (a zero of $f_0$) and expand
the time-dependent horizon radius around this as
\be
\label{exprbh}
r_h (T) = r_{h0} + x_h (T) 
\ee
where the shift in the position of the zero is assumed small, 
$|x_h | / r_{h0}   \ll 1$. $r_h(T)$ is defined as a zero of $f(r,T)$,
therefore $x_h$ can be deduced by Taylor expanding the relation
$f(T, r_h(T) )=0$ around $r_{h0}$ and using the expression we have 
derived for $f(r,T)$, yielding
\be\label{xbis}
x_h(T) = \frac{r_{h}}{2\kappa_h} \eta_0(r_{h}) 
\int_{T_0}^T\frac{\dps}{M_p^2} dT'
- \frac{I(r_h)}{2r_h} \frac{\dps}{M_p^2}
\ee
thus verifying \eqref{solextratwo}. 
Expanding $\eta$ near each horizon similarly shows
\be
\beal
\eta(r_h(T)) &= \eta_0(r_{h}) + x_h \eta_0'(r_h) 
+ \frac{r_h(1-\eta_0^2)}{2f_0} \int_{T_0}^T\frac{\dps}{M_p^2}
+\eta_1(r_h,T)\\
&= \eta_0(r_{h})+\eta_1(r_h,T) - \frac{I(r_h)}{2r_h}  \frac{\dps}{M_p^2}
\eeal
\ee
i.e.\ fixing $\eta_1 = I(r) \dps/2rM_p^2$, we have that 
$\eta(r_h(T)) = \eta_0(r_{h0}) = \pm 1$, explicitly confirming that
$g_{rr}$ remains regular at the new horizon radii.
 
Therefore although the metric function $g_{TT}$ vanishes at 
$r_{h}(T)$, the metric is non-degenerate there, as was the case in 
the unperturbed SdS spacetime.  The cross term $+2dTdr$ in the 
metric at the black hole horizon illustrates the ingoing 
nature of the $T$ coordinate near that horizon, where it becomes a 
null coordinate that is well-behaved\footnote{The vector 
$(\partial / \partial T)$ is null at $r=r_b$ and is normal to the surface 
$dr =0$, and the second null radial direction 
is defined by  $( -\eta_0'(r_b(T)) / \kappa _b )dr + 2dT =0$. 
So the two null directions on the black hole horizon are
\be\label{nullvectors}
T^a = \frac{\partial~}{ \partial T} \  , \quad 
K^a =  -\frac{\eta_0'(r_b(T))}{2\kappa _b}\frac{\partial ~}{ \partial T} 
- \frac{\partial~}{ \partial r}
\ee
normalized so that $T^a K_a = -1$.}, a property that was built into the 
choice of $T$ to ensure ingoing boundary conditions for $\phi (T)$.
A similar analysis applies to the cosmological horizon as $r$ approaches 
$r_c (T)$. The cross term in the metric in this case will have the opposite sign
illustrating the outgoing nature of the coordinate $T$ near the cosmological horizon.

The preceeding subsections are summarized by the formula for 
$r_h (T)$ given in  \eqref{solextratwo}, valid 
after an initial start-up time. 

\subsection{Calculation in null coordinates} 

In this subsection we show that these results for the growth of the 
horizon radii agree with the calculations in our previous paper 
\cite{Gregory:2017sor}. This is a useful exercise since the
first paper worked in null coordinates, in which it was simple to locate 
the horizons but difficult to find the metric throughout the region.
In contrast, the coordinates used in the current paper allow us to find 
the metric straightforwardly, but the horizon structure requires more work.
Comparison requires further processing of the results of 
\cite{Gregory:2017sor} using the slow-roll approximation requirements 
derived earlier. 

In \cite{Gregory:2017sor}, we showed that the linearized Einstein equations
on the horizon reduced to simple second order ODE's in terms of the advanced 
time coordinate $V_b$ on the black hole horizon, and $U_c$ on the cosmological
horizon:
\be\label{dotareas}
\frac{d^2~~}{dV_b^2}\delta {\cal A}_b = 
-\frac{{\cal A}_b}{\kappa_b^2M_p^2 }\frac{ \dot{\phi}^2}{V_b^2}\;,\qquad
\frac{d^2~~}{dU_c^2} \left ( U_c \delta {\cal A}_c \right)
= -\frac{{\cal A}_c}{\kappa_c^2 M_p^2}  \frac{ \dot{\phi}^2}{U_c}
\ee
where ${\cal A}_h$ is the area of the respective horizon.
In the background SdS metric, these coordinates are related to $T$ by
\be
\kappa_b V_b  = \exp [ \kappa_b (T-T_0) ]\;,\qquad
\kappa_c U_c = \exp[\kappa_c(T-T'_0) ]
\ee
Integrating \eqref{dotareas} gave the general expressions for the horizon
areas, from which we can deduce the values of horizon radii as
\be
\beal
\delta r_b &= -   \frac{ r_b }{ 6\gamma \kappa_b M_p^2 } V_b
\int_{V_b}^\infty \delta W[\phi(V_b')]  \frac{dV_b^\prime }{V_b^{\prime 2} }\\
\delta r_c &= - \frac{ r_c  }{6\gamma \kappa_c M_p^2 U_c} \int^0_{U_c}
\delta W[\phi(U_c^\prime )] dU_c^\prime
\eeal
\ee
At first sight, these look somewhat different from the values of $\delta r_h$
obtained from \eqref{solextratwo}, however, armed with the slow roll
analysis for integrating the Einstein equations, let us re-examine
these expressions. Integrating by parts in $\delta r_b$,
and using $\kappa_b V_b  = \exp [ \kappa_b (T-T_0) ]$ we find
\be
\delta r_b = -   \frac{ r_b }{ 6\gamma \kappa_b M_p^2  } \left [
\delta W[\phi(V_b)] -3\gamma e^{\kappa_b T }
\int_T^\infty  \dot{\phi}^2 e^{-\kappa_b T' } dT'\right] 
\ee
Similarly, the $r_c$ equation can be integrated by parts to give
\be
\delta r_c = - \frac{ r_c  }{6\gamma \kappa_c M_p^2 } \left [
- \delta W[\phi(U_c)] + 3\gamma e^{-\kappa_c T}
\int_T^\infty  \dot{\phi}^2 e^{\kappa_c T' } dT' \right]
\ee
To approximate these integrals, we use \eqref{slowdelta} to compare 
how rapidly $\dot{\phi}^2$ is varying compared to the exponential, 
in other words, the magnitude of
\be
\left | \frac{1}{\dot{\phi}^2 \kappa_h } \frac{d~}{dT} \dot{\phi}^2 \right |
= \left | \frac{W''}{\gamma \kappa_h} \right |
= \frac{r_c^2+r_cr_b + r_b^2}{r_c^2+r_b^2} 
\frac{4 r_h\delta }{2r_h + r_{\bar{h}}} < 3 \delta\ll1
\ee 
(where the `$\bar{h}$' subscript stands for `the other' horizon). 
Thus $\dot{\phi}^2$ varies much more slowly than the exponential, 
and we can approximate each of these integrals by 
$\dot{\phi}^2 e^{-|\kappa_h| T}/|\kappa_h|$ yielding
\be
\delta r_h \simeq  -   \frac{ r_h }{ 6\gamma |\kappa_h| M_p^2 } \left [
\delta W - \frac{3\gamma}{|\kappa_h|} \dot{\phi}^2 \right]
\label{deltahor}
\ee
Let us now compare these two terms. At the start of slow-roll,
i.e.\ at the local maximum of $W$, 
\be
W \simeq W_0 + \frac12 W'' \delta \phi^2 \;\; \Rightarrow
\quad \delta \phi \propto e^{-W'' T/3\gamma} \;\; {\rm as} \;\; T\to-\infty
\ee 
hence
\be
\left | \frac{3\gamma\dot{\phi}^2}{\kappa_h\delta W} \right |
= \left |\frac{2W''}{3\kappa_h\gamma}\right | < 2\delta \ll1
\ee
thus $\delta W$ is initially the dominant term in \eqref{deltahor}.
Moreover, throughout slow-roll, examining the rate of change of each term,
\be
\dot{\delta W} = -3 \gamma \dot{\phi}^2\;\;, \quad
\left ( \frac{3\gamma \dot{\phi}^2}{\kappa_h} \right )^{\!\boldsymbol{\cdot}}
= - 3 \gamma \dot{\phi}^2 \frac{2W''}{\kappa_h\gamma}\;\;\;\Rightarrow
\quad \left | \left ( \frac{3\gamma \dot{\phi}^2}{\kappa_h} 
\right )^{\!\boldsymbol{\cdot}} \right | \ll \left | \dot{\delta W} \right|
\ee
shows that if $\delta W$ is dominant initially, it will remain so throughout
the transition between vacua, meaning that the evolution of the horizon 
area is dominated by the shift in the cosmological constant which is the 
first term in \eqref{deltahor}, in agreement with \eqref{solextratwo}.

\section{Horizon areas, black hole mass, and first laws}
\label{horizonsandfirst}

In this section we compute the rates of change for the areas of the 
black hole and cosmological horizons, the black hole mass, and the 
thermodynamic volume between the horizons. We then show that the 
de Sitter patch first law \eqref{firstsds} and the mass first law 
\eqref{firstmasssds} are satisfied throughout the evolution. 
In our previous paper \cite{Gregory:2017sor} we showed that \eqref{firstsds}
held between the initial and final SdS states, relating the total changes in 
these quantities over the evolution. 
In  this earlier work we lacked the detailed form of the time dependent 
metric and were unable to verify the mass first law equation, though we 
inferred the value of the late time mass based on assuming that the mass 
first law was true. Here, equipped with the solution for the metric 
\eqref{perturbed}, we are able to do more.
 
\subsection{Horizon area growth and first law}

$A\ priori$ one expects that the black hole horizon area should be increasing 
in time due to accretion of scalar field stress-energy.  However, for the 
cosmological horizon there are competing influences on its area that 
tend in opposite directions. For fixed $\Lambda$ the cosmological horizon 
gets pulled in as the black hole mass grows, while for fixed black hole mass 
the cosmological horizon grows as $\Lambda$ shrinks. The solutions showed 
that both horizon radii are increasing,
so the latter effect dominates the behavior of the cosmological horizon.
 
Throughout this section we will work in the late time limit defined above in 
\S \ref{findingzeros}, so the black hole and cosmological horizons 
are located at the zeros of the quasi-static metric function $f_{QS}(r,T)$.
This implies that  $r_b (T)$ and $r_c (T)$ are given in terms of the time 
dependent mass $M(T)$ and vacuum energy $\Lambda (T)$ by the same 
relations \eqref{mlambda} that apply in the unperturbed SdS spacetime. 
The expressions for  $\dot{M}$ and $\dot{\Lambda}$ in the time dependent 
solution \eqref{perturbed} then yield the expressions given in \eqref{rhdot} 
for $\dot{r_b }$ and $\dot{r_c}$. It follows that  the 
rate of change of the black hole and cosmological horizon areas is
\be\label{dotarea}
\dot{A}_h = \frac{A_h}{|\kappa_h |} \frac{\dps}{M_p^2}  \  , \quad \quad h=(b,c)
\ee

Now it can be checked that the SdS-patch first law \eqref{firstsds} holds 
in a dynamical sense. Summing the expressions for the rate of growth 
of the horizon areas \eqref{dotarea} gives
\be\label{firstareas}
|\kappa_b| \dot{ A_b} +  |\kappa_c |\dot{ A_c} = A_{tot}\frac{\dps}{M_p^2}
\ee
Using the formulae for $\dot\Lambda$ in \eqref{derivs} and the 
thermodynamic volume of the de Sitter patch $V_{dS}$, equation 
\eqref{gammainterp}, gives
\be\label{firstvol}
V\dot \Lambda = -A_{tot}\frac{\dps}{M_p^2}
\ee
and so
\be\label{first}
|\kappa_b| \dot{ A_b} +  |\kappa_c |\dot{ A_c} = - V\dot \Lambda
\ee
 Interpreting the variations in \eqref{firstsds} as time derivatives, 
and translating $ |\kappa_h | M_p^2 \delta A_h =T_h\delta S_h$,  we 
see that the SdS-patch first law \eqref{firstsds} holds between 
successive times. Hence the decrease of the effective cosmological constant due to the scalar field rolling down 
its potential goes into increasing the black hole and cosmological horizon entropies.

It is also straightforward to check that the mass 
first law \eqref{firstmasssds} is satisfied throughout the evolution for the 
time dependent solutions. Plugging in for $\dot M$ and $\dot\Lambda$ 
using \eqref{perturbed}, and $\dot A_b$ using \eqref{dotarea}, one finds 
that the mass first law reduces to the first relation in \eqref{betagamma},
that is, the evolution satisfies
\be
\frac{\dot M}{M_p^2} - |\kappa_b|\dot A_b + V_b\dot\Lambda =0
\ee

The accumulated growth in time for each horizon is obtained by 
integrating the expressions \eqref{dotarea} for $\dot{A_h }$,
which at late times gives  
\be\label{changearea}
\delta A_h (T) 
= - \frac{A_h V_{dS}}{3 |\kappa _h | A_{tot} }   \delta\Lambda
\ee
where $\delta A_h (T) \equiv A_h (T) - A_{h0}$ and $\delta\Lambda 
=  \Lambda_0 -\Lambda [\phi (T)] $.
The prefactors all refer to values in the initial SdS spacetime, and 
keeping in mind that $\delta \Lambda(T)$ is negative,  $\delta A_h$ 
is positive. Hence the change in area of each horizon between $T_0$ 
and $T$ is proportional to the initial horizon area times the change in 
the effective cosmological constant.
It is also interesting to note that during the evolution
that the fractional increase in area, times the magnitude of the surface 
gravity, is the same for both horizons
\be\label{same}
|\kappa_c | \frac{ \delta A_c}{A_c} = |\kappa_b| \frac{\delta A_b}{A_b}
\ee

The geometrical quantities appearing in equation \eqref{changearea} 
are not all independent. The initial SdS spacetime is specified by two 
parameters, which we have been taking to be the initial black hole horizon 
radius and the initial value of the potential, and so one wants formulae that 
only depend on $r_b$ and $\Lambda $. Substituting in the surface gravity 
and thermodynamic volume gives, for the black hole horizon,
\be\label{deltaastep}
\delta A_b = A_b \frac{ |\delta \Lambda | }{ \Lambda} \  
\frac{ 2 r_b ( r_c ^2 + r_b^2 +r_c r_b )}{( 2r_b + r_c)(r_b^2 +  r_c^2 )}
\ee
The corresponding expression for $ \delta A_c$ is obtained by 
interchanging $r_b$ and $r_c$.  In \eqref{deltaastep} $r_c$ is still
an implicit function of $r_b$ and $\Lambda$. We will display the 
results in the limits of small and large black holes.
For the black hole horizon area one finds
\be\label{bhareachange}
\delta A_b \simeq  
\begin{cases}
A_b \frac{2  |\delta \Lambda |}{\sqrt{3}\Lambda}  \  
\sqrt{\Lambda r_b^2} \ ,\qquad\qquad & \sqrt{\Lambda r_b^2} \ll 1  \\
A_b \frac{ |\delta \Lambda |}{\Lambda} \ ,  
&   \sqrt{\Lambda r_b^2} \sim 1
\end{cases}
\ee
One sees that the fractional growth is parametrically suppressed for 
small black holes, and of order $| \delta \Lambda| /\Lambda $ for large 
ones. Likewise, one can examine how much the growth of the 
cosmological horizon is suppressed  by the presence of the black hole. 
One finds
\be\label{careachange}
\delta A_c \simeq 
\begin{cases}
12\pi \frac{ |\delta \Lambda |}{ \Lambda^2}  \ , \qquad  \qquad
& \sqrt{\Lambda r_b^2} \ll 1  \\
4\pi \frac{ |\delta \Lambda |}{ \Lambda^2}  \  ,   & \sqrt{\Lambda r_b^2} \sim 1
\end{cases}
\ee
The small black hole result is the same as if there were no black hole. 
On the other hand, the  growth of the area of the cosmological horizon 
can be diminished by as much as a factor of two-thirds for large black 
holes. This suggests that the effect of a large black hole on the spectrum 
of CMBR perturbations created during slow roll inflation is worthy of 
further study.

While both horizon areas increase, the black hole gets smaller compared 
to the cosmological horizon in certain ways. The volume $V_{dS}$  
between the horizons  increases according to
\be\label{changevol}
\dot V_{dS} =2\pi  \left( \frac{r_c^3}{|\kappa _c | } 
- \frac{r_b^3}{|\kappa _b| }\right) \frac{\dps}{M_p^2}
\ee
Since $|\kappa_c | < \kappa_b$ \cite{McInerney:2015xwa} the right hand 
side is positive. Another measure is how the difference between the black hole 
and cosmological horizon areas changes in time:
\be\label{areadiff}
A_c (T) - A_b (T) - (A_{c0} - A_{b0})=\delta A_c - \delta A_b 
= 24\pi \frac{ |\delta \Lambda (T) | }{\Lambda^2 } 
\frac{(r_c -r_b ) ( r_c +r_b )^3 }{(r_c^2 + r_b^2 ) (2r_c +r_b ) (2r_b +r_c ) }  
\ee
Since $|\delta \Lambda (T) | $ increases with time, the black hole is getting 
smaller in comparison to the cosmological horizon. To unravel the parameter 
dependence of \eqref{areadiff}, we again look at different limiting cases
\be\label{bhcosareadiff}
\delta A_c - \delta A_b   \simeq  
\begin{cases}
12\pi \frac{ |\delta \Lambda |}{ \Lambda^2 }  \   \sqrt{\Lambda r_b^2}  \ , \qquad
\quad\quad & \sqrt{\Lambda r_b^2} \ll 1  \\
6\pi \frac{ |\delta \Lambda |}{\Lambda^2 } \ , & r_b= {1\over 2} r_c \\
\frac{32 \pi}{3} \frac{ |\delta \Lambda |}{ \Lambda^2 } (1- \sqrt{\Lambda r_b^2} )  \ \  , 
& \sqrt{\Lambda r_b^2} \sim 1
\end{cases}
\ee
We see in this case that the effect is parametrically suppressed both 
for very small and very large black holes and most prominent in the 
intermediate regime.
 
\subsection{Dynamical temperature}

We have not yet discussed horizon temperature for the time dependent 
quasi-SdS black holes \eqref{perturbed}. These represent 
non-equilibrium systems, for which it is not clear that a well-defined 
notion of dynamical temperature should exist. Nevertheless, given that our 
system is only slowly varying, we might expect that an adiabatic notion 
of temperature makes sense \cite{GalvezGhersi:2011tx,Barcelo:2010pj}
and indeed candidate definitions have 
been suggested in the literature.  We will focus, in particular, on the 
proposal \cite{Hayward:2008jq} which defines a dynamical surface gravity 
$\kappa_{dyn}$ for outer trapping horizons in nonstationary, spherically 
symmetric spacetimes.  In this construction, the Kodama vector 
\cite{Kodama:1979vn} substitutes for the time translation Killing vector 
of a stationary spacetime in providing a preferred flow of time.  
Further, the authors use a variant of the tunneling method of 
\cite{Parikh:1999mf}, adapted to the  non-stationary setting, to 
argue that particle production has a thermal form with 
temperature $T_{dyn}=\kappa_{dyn}/2\pi$.

The dynamical surface gravity is defined in \cite{Hayward:2008jq} as
\be\label{genkappa}
\kappa_{dyn} = \frac{1}{2}\star d \star d r
\ee
where the Hodge $\star$ refers to the two dimensional $T,r$ subspace
orthogonal to the $2$-sphere, 
and the quantity is to be evaluated at the horizon.
The Hodge duals of the relevant forms are given by
\be
\star d r = f dT - h dr \  , \qquad \star dr \wedge d T= 1
\ee
Using these expressions, and evaluating \eqref{genkappa} on the black hole 
horizon, one finds that
\be\label{tempone}
\kappa_{dyn} (T)   = (f'+\dot{\eta} )
= \kappa_b (T) + {\cal O}(\varepsilon\delta)
\ee
where $\kappa_b(T)$ was defined in equation \eqref{kbdef} as the 
derivative of the perturbed metric function $f(r,T)$ evaluated at
radius  $r_b (T)$. Note that although the contribution $\delta f(r,t)$ to 
the metric function vanishes at $r_{b0}$, its derivative is non-zero and 
therefore $\kappa_b(T)$ gets contributions both from $f_{QS}$ and 
from $\delta f$, with the result that
\be\label{derivf}
\kappa_b (T) 
= \kappa_{QS}(T)  -\frac{C_b }{2\kappa_b} \dps
\ee
Here the quasi-static surface gravity $\kappa_{QS}(T)$ is given by the 
SdS relation \eqref{kappah} using $r_b (T)$ and $r_c (T)$,
and expanding to linear order in the perturbative quantities. 
These time-dependent corrections to the radii $r_h (T)$ are proportional
to the integral over time of $\dps$, but the second term in \eqref{derivf} 
depends on the instantaneous value of $\dps$. So after an initial time 
period the latter term is small compared to the first, and the dynamical
surface gravity is well approximated by
\be\label{latekappa}
\kappa_b (T)\simeq \kappa_{QS} (T)
\ee
We see that the proposal of \cite{Hayward:2008jq} yields a simple, intuitive result.
This is in agreement with our results in \cite{Gregory:2017sor} where the calculation
was done using null coordinates.

\section{Conclusion}
\label{conclusion}

In this paper we have found a tractable form for the metric of a black 
hole in a slow-roll inflationary cosmology, to first order in perturbation 
theory, which one can readily understand in terms of expectations for 
the slowly evolving system. The solution directly gives the time dependence 
of $\Lambda$ and $M$ and it is straightforward to then find the time dependent
horizon areas, the thermodynamic volume, and the dynamical surface gravity. 
A topic for future study is to compute the flux of the energy-momentum 
of the scalar field across surfaces of constant $r$.
The flux is ingoing at the black hole horizon, and outgoing at the cosmological 
horizon, and it would be good to understand our results for the growth of the 
horizons and the mass in terms of the fluxes in greater detail,
as was done in \cite{Chadburn:2013mta}.
Further, there must be a transition surface between the horizons where the 
flux vanishes, so mapping out the flux
throughout the domain would be of interest. A related issue is to look at the 
energy density and pressure variation across spatial slices that interpolate 
to a standard inflationary cosmology in the far field. The range of validity of 
the approximate solution we have derived is another issue for further study. 
Conservatively, one assumes that $| \delta  \Lambda / \Lambda|$
must be small. However, since the slow-roll conditions require that the 
derivatives of the potential must be small, one might ask if larger accumulated 
change in $W$ is allowed, as long as the evolution is slow enough.
In any case, our approximation is equally valid as the slow roll
approximation in the inflationary evolution of the early universe.

It is of interest to calculate the perturbations from inflation with the black hole 
present, as such signatures in the CMBR may be a method for detecting, or 
inferring, primordial black holes \cite{Afshordi:2017use}.
Although the black hole itself is small-scale, 
the wavelength of its signature on modes that re-enter the horizon at late times 
is stretched with the modes themselves. A related problem is to compute the 
Hawking radiation in this metric by extending the methods of
\cite{Kastor:1993mj} to the quasi-static case. Such a calculation is needed to 
support the interpretation of the dynamical surface gravity as a physical temperature. 
While this quasi-static set-up would only represent a step in understanding 
horizon temperature in a dynamical setting, this metric is one of the few
known dynamical  examples where the cosmological and black hole 
temperatures are not equal.

There are examples of  elegant analytic descriptions in which 
an evolving physical system tracks a family of static solutions, 
such as charge-equal-to mass black holes 
\cite{Hartle:1972ya,Ferrell:1987gf,Traschen:1992wy}, or magnetic monopoles
\cite{Montonen:1977sn, Atiyah:1985dv}, with small relative velocities. 
In these cases there is a BPS symmetry of the zeroth order time-independent 
solution, which apparently protects small perturbations from being too disruptive. 
There is not obviously any such symmetry in the black hole plus scalar field 
system, and yet the slow-roll dynamics is analogous. One avenue for future 
study is to see if there is an underlying reason for this behavior, which in turn 
could lead to a more fundamental understanding.

\begin{acknowledgments}

RG is supported in part by the Leverhulme Trust, by
STFC (Consolidated Grant ST/P000371/1),
and by the Perimeter Institute for Theoretical Physics.
Research at Perimeter Institute is supported by the Government of
Canada through the Department of Innovation, Science and Economic 
Development Canada and by the Province of Ontario through the
Ministry of Research, Innovation and Science. 

\end{acknowledgments}

\providecommand{\href}[2]{#2}


\end{document}